\documentstyle[twocolumn,aps,prl]{revtex}
\begin{document}

\draft

\twocolumn[\hsize\textwidth\columnwidth\hsize\csname @twocolumnfalse\endcsname

\title{Charging induced asymmetry in molecular conductors}

\bigskip

\author{F. Zahid, A. W. Ghosh, M. Paulsson, E. Polizzi and S. Datta}
\address{ School of Electrical and Computer Engineering, Purdue
University, W. Lafayette, IN 47907}%

\maketitle

\medskip

\widetext
\begin{abstract}
We investigate the origin of asymmetry in various
measured current-voltage (I-V) characteristics of molecules
with no inherent spatial asymmetry, with particular focus on
a recent break junction measurement.  We argue that such
asymmetry arises due to unequal coupling with the contacts
and a consequent difference in charging effects, which can
only be captured in a self-consistent model for molecular
conduction.  The direction of the asymmetry depends on the
sign of the majority carriers in the molecule.  For
conduction through highest occupied molecular orbitals (i.e.
HOMO or p-type conduction), the current is smaller for
positive voltage on the stronger contact, while for
conduction through lowest unoccupied molecular orbitals
(i.e.  LUMO or n-type conduction), the sense of the
asymmetry is reversed.  Within an extended H\"uckel
description of the molecular chemistry and the contact
microstructure (with two adjustable parameters, the
position of the Fermi energy and the sulphur-gold bond length), an appropriate description of
Poisson's equation, and a self-consistently coupled
non-equilibrium Green's function (NEGF) description of transport, we
achieve good agreement between theoretical and
experimental I-V characteristics, both in shape as well as
overall magnitude.
\end{abstract}
\bigskip

%31.15.Ar Ab initio calculations
%81.07.Nb Molecular Nanostructures
%81.07.Lk Nanocontacts
%85.65.+h Molecular electronic devices
%72.10.Bg General formulation of transport theory
%72.20.Dp General theory, scattering mechanisms of conductivity
%73.23.-b Electronic transport in mesoscopic systems
%73.40.Sx Metal-semiconductor-metal structures
%73.63.-b Electronic transport in mesoscopic or nanoscale materials and structures
%2col
%end of wide text
]
\narrowtext
%2col

Future electronic devices are quite likely to incorporate
molecular components, motivated by their
natural size, mechanical flexibility and chemical
tunability.  Encouraging progress in this direction has been
achieved with the capacity to self-assemble, functionalize
and reproducibly measure the current-voltage (I-V)
characteristics of small groups of molecules.  Molecular
I-Vs reveal a wide range of conducting properties,
from metallic conduction in carbon nanotubes \cite{rNT} and quantum
point contacts \cite{rQPC}, to semiconducting behavior in DNA \cite{rDNA}
and conjugated aromatic thiols \cite{rReed1}, and insulating behavior
in alkylthiol chains \cite{ralkane}.  Interesting device
characteristics such as rectification \cite{rMetzger}, switching \cite{rReed2} and
negative differential resistance on silicon substrate \cite{rHersam} have
also been reported.  In particular, molecular rectification
continues to be a widely studied property, right from the
earliest days of this field.

The classic paradigm for asymmetry in molecular I-V
measurements is the Aviram-Ratner diode, consisting of a
semi-insulating molecular species bridging an electron
donor-acceptor pair \cite{rAviram}.  A positive bias on the contact at
the donor end brings the energy levels on the donor and
acceptor sites into resonance, while the opposite bias moves
the system away from resonance, leading to a strongly
asymmetric I-V characteristic \cite{rMetzger}.  
Another example of asymmetric I-V is when the electrostatics of the system is
dominated by one contact, e.g., a gated molecule where the gate is electrically connected to
the source electrode \cite{rtrans,rLiang}. This gives rise to a spatially asymmetric
electrostatic potential over the molecule which in turn generates a strong 
asymmetry in the I-V. In these, as well as
most commonly studied examples of rectification, some kind of spatial
asymmetry in the system seems essential, causing the
energy levels, the electrostatic potential and the electron
wave functions to be quite different for positive and
negative voltages \cite{rVuillame}. Typically, such asymmetry leads to peaks of {\it{similar heights}} 
in the conductance-voltage (G-V) characteristic occurring at
{\it{different bias values}}.

In this paper, we address an I-V asymmetry observed for a
{\it{spatially symmetric}} molecule \cite{rReichert,rKerg,rReifenberger} that is qualitatively
different from and weaker than the rectifications in the above mentioned
situations.  The molecular I-V curves for these
systems start off being symmetric, but pick up a weak,
reversible asymmetry as the contacts are manipulated.  In
contrast to the two cases of rectification mentioned above, conduction in
these molecules at opposite voltages occurs essentially
through the same molecular levels with very similar wave
functions.  For resonant conduction in particular, this
asymmetry shows up as conductance peaks of {\it{different heights
occurring at symmetrically disposed voltage values}}.  We show
that the origin of the observed contact-induced asymmetry is
nontrivial, and involves self-consistent shifts in the
energy levels due to charging effects.  Asymmetry in
charging arises due to unequal coupling with the contacts,
and seems to be present in conduction measurements performed with
a break junction \cite{rReichert} or an STM tip \cite{rReifenberger}. 

Although this work addresses the origin of a weak
rectification effect that may or may not be of practical
significance from a device point of view, it serves three additional purposes:  
(a) using a computationally inexpensive yet rigorous
self-consistent transport model we 
show the significance of charging effects in explaining experimentally
observed I-V characteristics; (b) our
calculated I-V (Fig. 1) is not just qualitatively, but also
quantitatively in agreement with break-junction measurements
on very small groups of molecules \cite{rReichert}.  This is a significant
deviation from the typical situations where theoretical estimates of current values 
differ from experimental currents by orders of magnitude \cite{rLang,rDamle,rStok,rEmber}; (c) the
physics also sheds light on the nature of the charge
carriers, i.e.  whether conduction is n-type or p-type.  The
identification of the polarity of charge carriers is of obvious
importance in semiconductor devices, since electrons and
holes have different effective masses, leading to different
mobilities.  By analogy, molecular LUMO and HOMO levels have
quite different wavefunctions \cite{rZahid}, leading to different
transmissions and current conducting properties.  Much
uncertainty exists about the nature of the conduction
orbitals, or equivalently, the position of the Fermi energy
relative to the molecular energy levels \cite{rDamle,rLang,rEmber,rHall,rghosha}.  
We argue that for
the kind of contact-induced asymmetry discussed here,
current is lower for positive bias on the strongly coupled
contact for HOMO conduction, and higher if conduction is
through a LUMO level.

A description of molecular conduction can be put on a
quantitative footing by using an appropriate molecular Fock
matrix $F$ (ab-initio or semi-empirical), coupled with a non-equilibrium
Green's function (NEGF) formulation of transport \cite{rDamle,rDatta1}.
For a given Fock matrix $F$, overlap matrix $S$, and contact
self-energies $\Sigma_{1,2}$ with corresponding broadenings
$\Gamma_{1,2} = i\left(\Sigma_{1,2}-\Sigma_{1,2}^\dagger\right)$, the
energy levels are given by the poles of the nonequilibrium Green's
function $G$, while their occupancies are obtained from the
corresponding density matrix $\rho$, the contact electrochemical potentials $\mu_{1,2}$ 
and the Fermi functions $f_{1,2}$:
\begin{eqnarray}
G(E) &=& \left(ES-F-\Sigma_1-\Sigma_2\right)^{-1}\nonumber\\
\rho &=& \left(1/2\pi\right)\int_{-\infty}^{\infty}dE\left(f_1G\Gamma_1 G^\dagger + 
f_2G\Gamma_2 G^\dagger\right)\nonumber\\
f_{1,2}(E) &=& \left[ 1 + \exp{((E - \mu_{1,2})/k_BT)} \right]^{-1}
\label{e1}
\end{eqnarray}
The number of electrons $N$ and the steady-state current $I$ are then given by:
\begin{eqnarray}
N &=& 2~(\rm{for}~\rm{spin}) \times {\rm{trace}}(\rho~S) \nonumber\\
I &=& {{2e}\over{h}}\int_{-\infty}^\infty dE~ T(E,V)\left[f_1(E)
- f_2(E)\right]
\label{e2}
\end{eqnarray}
  where the transmission $T$ is given by:
\begin{eqnarray}  
T(E,V) &=& {\rm{trace}} (\Gamma_1 G \Gamma_2 G^\dagger)
\label{e3}
\end{eqnarray}
  
It has been pointed out that the nature of the molecular I-V
could sensitively depend on the self-consistent potential
profile \cite{rDatta2}.  
%It is imperative therefore to include charging
%effects self-consistently in our transport model in order to
%explain the asymmetry in I-V that arises from asymmetric
%coupling for a symmetric molecule.  
In our calculations, the
self-consistent charging effect is included within an
extended H\"uckel (EHT) \cite{rDatta2,rZahid} description of the Fock matrix
F, coupled with the NEGF equations described above.  The
self energies are calculated for Au (111) contacts in EHT
using a recursive technique \cite{rZahid}.  Both charging and
screening effects are incorporated in F through a
self-consistent potential $U_{SC} = V(\rho)$ which describes the
potential profile of the molecule under applied bias:
\begin{equation}
V(\Delta \rho) = V_{Laplace} + \Delta V_{Poisson}(\Delta \rho) + \Delta V_{image}(\Delta \rho)
\end{equation}
where $\Delta \rho$ represents the change in density matrix under bias, $\rho - \rho_{eq}$.
Equations (1) and (4) are then solved self-consistently to obtain the converged Fock matrix F. 
Any asymmetry in the potential profile
is included explicitly in the self-consistent potential $U_{SC}$.

\begin{figure}
\vspace{3.0in}
\hskip -0.5cm\includegraphics{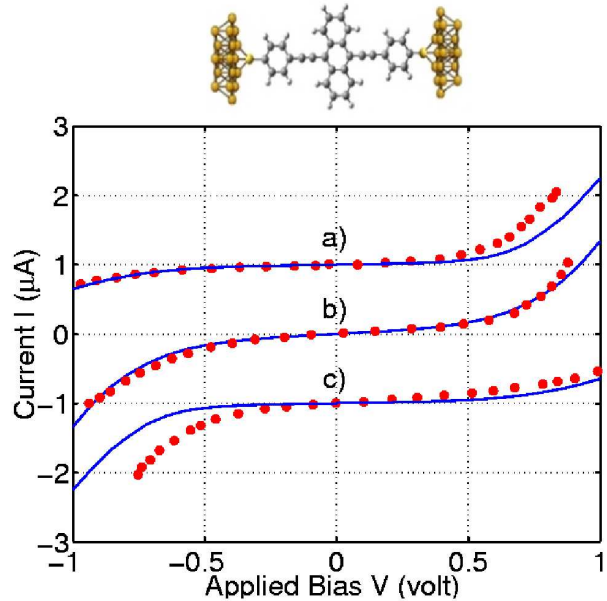}
\caption{(Color online) I-V characteristics of the gold-molecule-gold system shown above.
Solid line: theoretical calculations; dots: experimentally
obtained data in a break junction configuration \protect\cite{rReichert}. 
The molecular energy levels are raised by a constant potential $V_c$ (treated as a fitting parameter)
relative to the gold Fermi energy (-9.5 eV) in order to simulate the effect of charge 
transfer and band line-up at equilibrium. 
The same value of $V_c$ (1.5 eV) is used to generate 
all three curves, although allowing slight variation gives a better fit.
The upper curve (a) is obtained on decreasing the left 
electrode coupling by stretching the sulphur-gold bond length from its
equilibrium value of 2.53 $\mathring{A}$ to 3.18 $\mathring{A}$ and the lower curve (c) corresponds to 
the reverse, whereas curve (b) represents the symmetric coupling situation.
The bias polarity is defined as positive when the applied voltage on the left contact
is positive. Curves (a) and (c) are offset by +1$\mu$A and -1$\mu$A respectively for better visibility.} 
\label{f1} 
\end{figure}

The Poisson part in equation (4) is approximated using the
CNDO (complete neglect of differential overlap) method \cite{rPople,rMurell}, of which
only the Hartree potential is being utilized in our
treatment.  Both charging and screening effects 
are incorporated into this term. In CNDO approximation, this Poisson term becomes:

\begin{eqnarray}
\Delta V_{\rm{Poisson}}^M(\Delta\rho) &=& (\Delta\rho)_M\gamma_M + \sum_A^{~~~~~\prime} (\Delta\rho)_A
\gamma_{MA} \nonumber\\
where
~(\Delta\rho)_M &=& \sum_M(\Delta\rho)_{\mu\mu}^M \nonumber\\
\gamma^{}_M &=& \left( \mu_M^2 | \mu_M^2\right) \nonumber\\
\gamma^{}_{MA} &=&  \left( \mu_M^2 | \mu_A^2\right) \nonumber\\
M,A &=& Atomic~sites\nonumber\\
\mu &=& Slater~type~atomic~orbital
\end{eqnarray}
The notations used in Equation (5) are consistent with \cite{rMurell}. 
The two electron integrals $\gamma$'s, are the CNDO parameters which are obtained
from the experimental data and empirical fitting. 
These CNDO parameters allow us to capture the main physical
characteristics of the e-e interaction without evaluating the computationally expensive
two electron integrals directly.

\begin{figure}
\vspace{2.4in}
\hskip -0.6cm\includegraphics{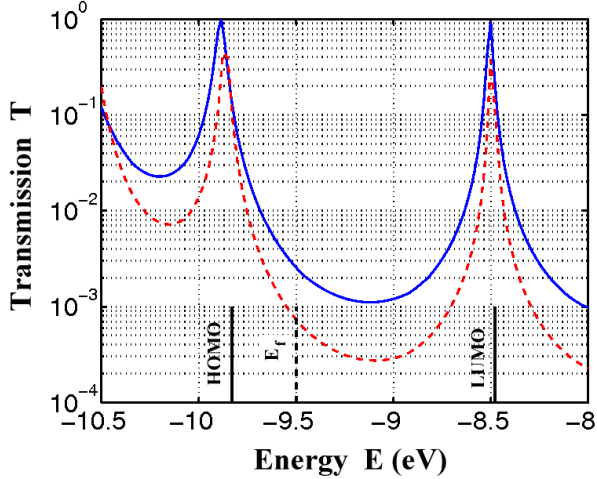}
\caption{(Color online) Equilibrium transmission coefficient as a function of energy at equilibrium.
Solid line: symmetric coupling; dotted line: asymmetric coupling with the left contact weakly
coupled. The transmission coefficients do not show any significant difference in the energy
range of interest from -10 eV to -9 eV. As current is proportional to transmission, this
explains why current values do not differ appreciably although the S-Au bond length has been
stretched by 0.65 $\mathring{A}$ in the asymmetric coupling situation.} 
\label{f2} 
\end{figure}

The Laplace and image potentials are calculated
by using a finite element method \cite{rPolizzi} treating the atomic sites
as points in free space. The two electrodes are treated as 
metallic plates [100 $\mathring{A}$ $\times$ 100 $\mathring{A}$] separated by the molecular
length. The molecule is placed in between the two plates. The Laplace
part is then obtained by solving Laplace's equation in 3D with the boundary conditions
set by $\mu_{1,2} = E_f \mp 0.5V_{appl}$ ($V_{appl}$: applied bias). The image part
is calculated similarly but with different boundary conditions
set by the potentials on the metallic plates due to point charges on the atomic
sites. The inclusion of image potential in our model does not make any significant
differences other than lowering the charging energy of the system by around 0.1 eV.

To illustrate charging induced I-V asymmetry
we apply our self-consistent transport model to the symmetric molecule, 
[9,10-Bis(($2^{'}$-para-mercaptophenyl)-ethinyl)-anthracene] (see top of
Fig. 1) studied in a recent break junction measurement \cite{rReichert}.   
The curve (b) in Fig. 1 is obtained assuming ideal sulphur-gold bonding (2.53 $\mathring{A}$)
on both sides of the molecule (symmetric
coupling situation) on a gold-molecule-gold system.  On the
other hand, the curves (a) and (c) represent an asymmetric
coupling situation where the asymmetry is introduced in the
system by stretching the bond between sulphur and gold on the left side 
and the right side respectively by 0.65 $\mathring{A}$. 
The sulphur-gold bond length has been treated as a fitting parameter which is
justifiable as the exact bond length is not known from the experiment. 

\begin{figure}
\vspace{1.8in}
\hskip -0.2cm\includegraphics{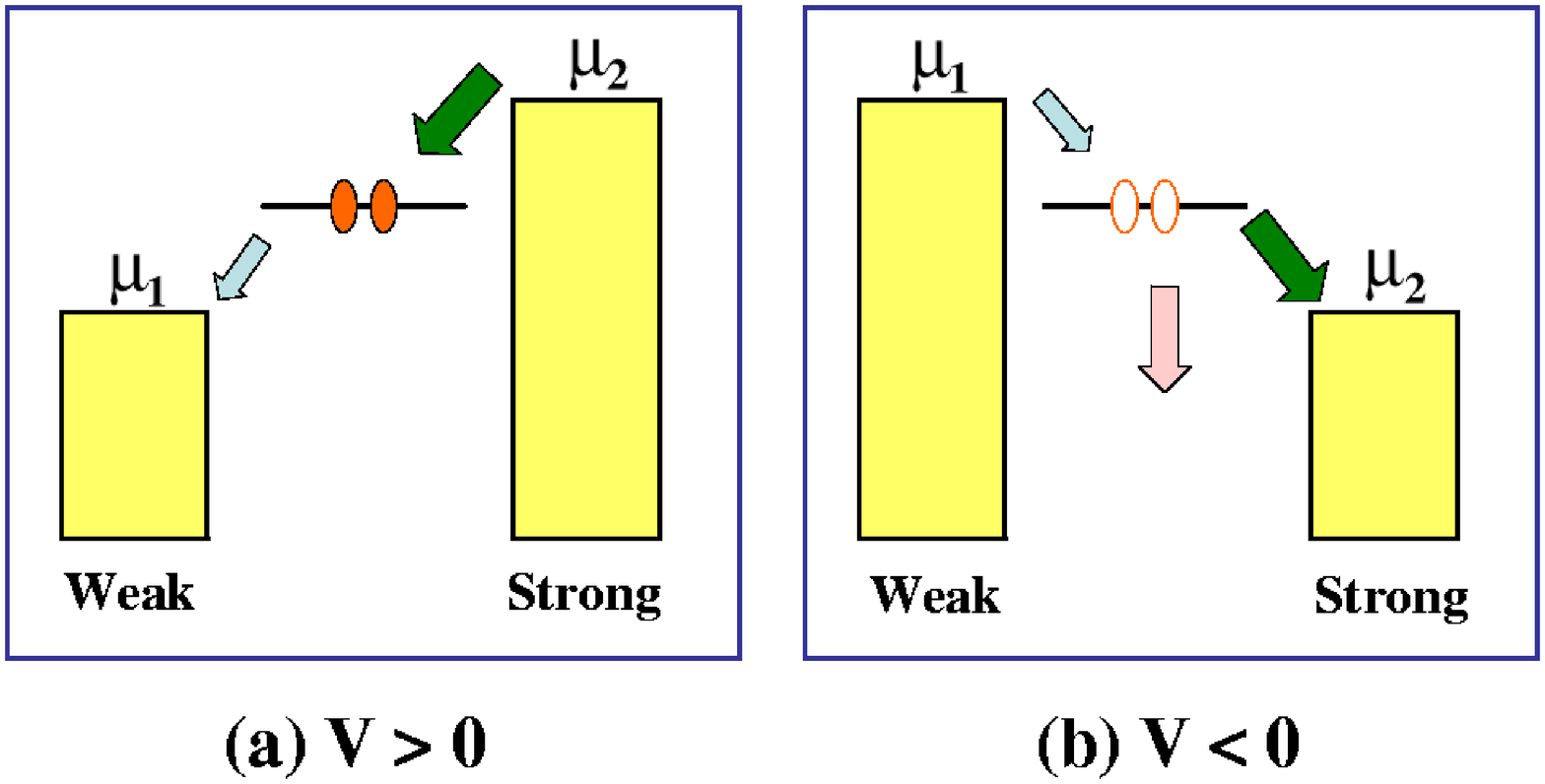}
\caption{(Color online) Origin of asymmetry due to charging shown for an one level
model: one side (say the right substrate) is strongly coupled. 
Although the same level is crossed by the contact
electrochemical potentials ($\mu_1$ and $\mu_2$) in both bias directions, for negative applied bias (b)
the HOMO level is emptied out by the strongly coupled
contact, which positively charges the molecule and shifts
the energy levels down.  Such a shift, not present for
positive applied bias (a), postpones the onset of conduction and
effectively stretches out the voltage axis in the I-V along the direction
of negative applied bias.} 
\label{f3} 
\end{figure}

Another fitting parameter used in our calculation is a constant potential 
$V_c$ which is applied on the molecule (only on the whole molecule and not on any part
of the gold contacts) in order to move the molecular 
energy levels up or down rigidly relative to the contact Fermi energy.
The precise location of equilibrium Fermi energy $E_f$ can depend sensitively
on many factors such as surface conditions, environmental situations, geometrical fluctuations etc.
In the absence of detailed information of all these factors in a particular experimental
set-up, it is {\it{necessary}} to treat the position of $E_f$ as an adjustable parameter.
Thus we believe it is justifiable to include this adjustable parameter $V_c$ in our calculations
to take into account of the effect of chrage transfer at equilibriumin and
adjust the position of $E_f$ relative to the molecular energy levels.
Our fitting requires that $E_f$ be closer to HOMO rather than LUMO 
i.e. the conduction is p-type. The Fermi energy of gold (within the EHT description) is
kept fixed at -9.5 eV \cite{rZahid} and relative to this value the molecular energy 
levels are moved up by $V_c = 1.5 eV$ so that $E_f-E_{HOMO}$ is set to be 0.33 eV.

\begin{figure}
\vspace{4.2in}
\hskip -0.8cm\includegraphics{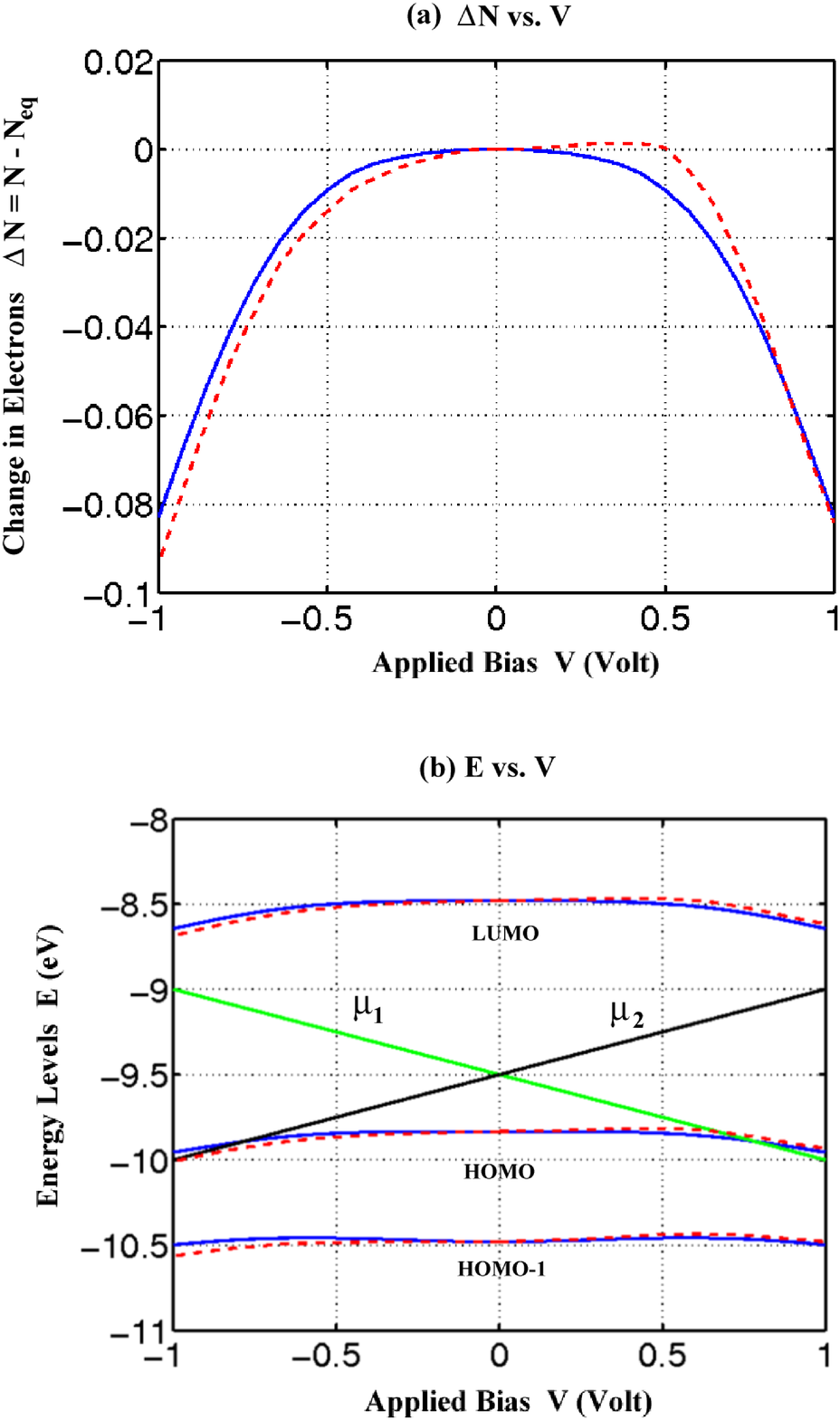}
\caption{(Color online) (a) Change in the number of electrons and (b) the molecular energy levels
as a function of applied bias. Solid line: symmetric coupling; dotted line: asymmetric 
coupling with the left contact weakly coupled. Both the number of electrons and the energy levels change
symmetrically with applied bias in the symmetric coupling situation, as expected. On the other hand,
for the asymmetric coupling case more electrons empty out in the negative bias 
direction (a) making the charging asymmetric and moving down the energy levels more
in the corresponding bias direction (b).} 
\label{f4} 
\end{figure}

The calculated trend in the I-V characteristics agrees
both in shape and current magnitude
with the experimental result, as evident from Fig. 1. The agreement of our
calculated current value with experiment suggests that the
observed I-V is obtained for a near ideal gold-molecule-gold
contact and the current is going through a single or at
most a few number of molecules. In spite of stretching the S-Au bond by 
0.65 $\mathring{A}$ the amplitude of current does not change appreciably. This can be 
explained by noting that the transmission coefficients do not differ significantly
in the energy range of interest from -10 eV to -9 eV (see Fig. 2). 
The little discrepancy in the shape of the
I-V can be attributed to the variation in the experimental results. It should also be
noted that the same value for the constant potential $V_c$ (i.e. same $E_f-E_{HOMO}$) 
is used to generate all the three curves in Fig. 1,
though the position of the Fermi energy can conceivably be different for 
curves (a) and (c) as the experimental conditions are different.
Indeed, by using slightly different
values for $V_c$ within a justifiable range ( $V_c$ = 1.55 eV and $V_c$ = 1.6 eV 
for curves (a) and (c) 
respectively) we do obtain a better fit of the I-V (not shown). 

In our calculations we assumed that the system is in the self-consistent field regime.
This assumption should be valid as long as the energy level broadening due to contact coupling
is comparable to (or greater than) the single-electron charging energy.
We estimate these factors to be of the order of 0.2 eV and 1.5 eV, respectively.
It should be further noted that the experimental current value is higher (of the order of 
$\mu$A) than other typical experimental values, suggesting strong chemisorbed bonding
between sulphur and gold at both ends. Thus we believe our assumption of self-consistent
field approach is a correct one.

\begin{figure}
\vspace{3.4in} 
\hskip -1.0cm\includegraphics{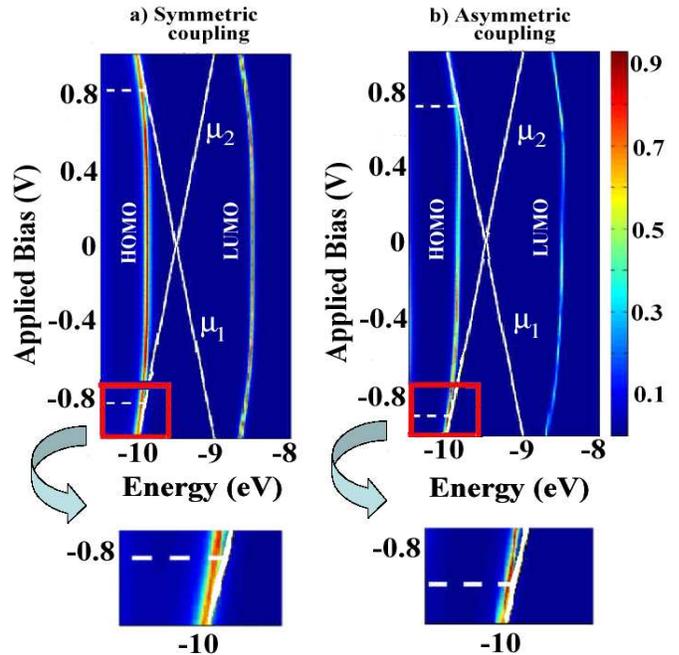}
\caption{(Color online) A color plot of the transmission as a function of energy
and applied bias. 
Fig. 5a
corresponds to symmetric coupling situation (Fig. 1b) whereas Fig.
5b corresponds to the asymmetric coupling situation where
the left side is weakly coupled (Fig. 1a). The white dotted lines
pinpoint the onset of level crossing in both bias directions (enlarged portions shown below). 
The Fermi energy of the device (-9.5 eV) is assumed to be closer to the HOMO level and
only this level is conducting in the applied bias range.}
\label{f5} 
\end{figure}

In order to explain the origin of this asymmetry in the I-V
we need to understand the effect of asymmetric
charging on molecular conduction.
This is explained
schematically in Fig. 3 for a simple single level model. Let us suppose that the
contacts are strongly asymmetric ($\Gamma_2 \gg \Gamma_1$). For
positive bias on the weak contact (Fig. 3a), the latter is trying to empty
the nearest (HOMO) level, while the strong contact is trying to fill
it, with the net result that the HOMO level stays filled. Current
onset occurs at the voltage where $\mu_1$ first crosses the (neutral)
molecular level.  For opposite bias direction (Fig. 3b), however, the HOMO level is
emptied out, which charges up the molecule positively. This adds a
self-consistent charging energy that lowers the energy level,
postponing thereby the point where the HOMO is crossed by $\mu_2$. In
effect, this stretches out the voltage axis in the I-V, leading thus to a smaller
conductance for negative bias on the weak contact.  For
LUMO-based conduction the argument is reversed, since filling the LUMO
level charges up the molecule negatively.

It can be seen from Fig. 4 that the number of electrons and the energy levels of 
the molecule  change symmetrically with the applied bias for the symmetric coupling
situation. This is expected for a symmetric system. However, in the asymmetric coupling
situation the weakly coupled left contact can not fill up the HOMO level quickly enough
and so this level remains more empty in the negative bias direction compared to that in
the positive bias (Fig. 4a). This in turn makes the charging asymmetric and forces the
energy levels to move down more in the negative bias direction (Fig. 4b). 

Fig. 5 shows a color plot of the transmission T(E,V), for symmetric
coupling (a) and asymmetric coupling (b), corresponding
respectively to curves (b) and (a) in Fig. 1.  In both cases, the same HOMO
level is crossed by the contact electrochemical potentials $\mu_{1,2}$ (white solid
lines) leading to current conduction. For symmetrically coupled contacts,
current starts flowing around $\pm 0.8$ V in both bias directions, leading to the
symmetric I-V in Fig. 1b.
For the asymmetrically
coupled case, however, differential charging leads to a postponed onset
of conduction at around -0.9 V for negative bias on the weak (left)
contact, and an earlier onset at around +0.7 V for opposite bias
polarity. This makes the current lower and stretched out in the negative bias 
direction which leads to the asymmetric I-V in Fig. 1a. 
The delay in the onset is seen clearer in the enlarged portions 
shown at the bottom of Fig. 5.

The charging induced asymmetry in the I-V also sheds light on whether the
current conduction is p-type or n-type. If we would
assume LUMO conduction in our calculations, the direction of asymmetry in the I-V
would get reversed.  Thus, from the knowledge of the coupling asymmetry and the I-V asymmetry
we can predict whether it is HOMO or LUMO conduction. 
Although we do not know which contact is weaker 
for the break junction measurements, for similar measurements with STM \cite{rReifenberger}
where the weaker contact is easily identified, this physics helps infer that 
conduction is p-type.
Most often it is very difficult to pinpoint
the position of Fermi energy relative to the molecular energy
levels and so this ability of predicting conduction type is
very desirable and useful. 
It should be noted that the nature of conduction could possibly be identified by other 
techniques, e.g. by using thermoelectric \cite{rMagnus} or gating effects
\cite{rKagan}.

It is worthwhile pointing out that the I-V asymmetry discussed in this paper 
{\it{can not}} be explained
solely in terms of asymmetry in the Laplace potential, neglecting
charging effects altogether. Asymmetric charging gives conductance peaks of 
{\it{different heights at
same voltage values}} (see Fig. 10 in \cite{rReifenberger}), in contrast to Laplace 
potential asymmetries which give conductance peaks at 
{\it{different voltages but of the same height}}. 

In summary, we have achieved a quantitative agreement with 
experimentally observed asymmetric I-Vs and established that
even for spatially symmetric molecules, contact asymmetry
can induce an asymmetric I-V through asymmetric charging.
The sense of the asymmetry depends on whether conduction is
through a HOMO or a LUMO level.

We would like to thank P. Damle, R. Reifenberger, D. Janes and H. B. Weber
for useful discussions. This work has been supported by NSF Network for
Computational Nanotechnology and the
US Army Research Office (ARO) under grants number EEC-0228390 and
DAAD19-99-1-0198.

\end{document}